\documentclass[runningheads]{llncs}
\usepackage[T1]{fontenc}
\usepackage{graphicx}
\usepackage{subcaption}
\usepackage{lipsum}
\usepackage{bm}
\usepackage{amssymb}
\usepackage{amsmath}
\usepackage{multirow}
\usepackage{pifont}
\usepackage{hyperref}

\usepackage[]{acronym}
\acrodef{AI}{Artificial Intelligence}
\acrodef{DL}{Deep Learning}
\acrodef{US}{Ultrasound}
\acrodef{PCCL}{{\bf P}ixel-level and {\bf C}lass-level {\bf C}onsistency {\bf L}earning}
\acrodef{CNN}{Convolutional Neural Network}
\acrodef{ROI}{Region of Interest}
\acrodef{SSL}{Semi-Supervised Learning}
\acrodef{AL}{Active Learning}
\acrodef{GT}{Ground Truth}
\acrodef{CR}{Consistency Regularization}
\acrodef{EMA}{Exponential Moving Average}
\acrodef{MAC}{Mutual Agreement Consistency}
\acrodef{KL}{Kullback-Leibler divergence}
\acrodef{MIG}{Mutual Information Gap}
\acrodef{CE}{Cross-Entropy}
\acrodef{Dice}{Dice based coefficient}

\newcommand{\cmark}{\ding{51}} 
\newcommand{\xmark}{\ding{55}} 

\begin{document}
\title{A Clinician-Centered Pipeline for Annotation and Evaluation in Ultrasound AI Studies}
%
%
\author{
Fangyijie Wang\inst{1,2,\dagger}\orcidID{0009-0003-0427-368X} \and
Jianjun Yu\inst{3,\dagger} \and
Wentao Shi\inst{4} \and
Haixia Huang\inst{5} \and
Ran Shi\inst{4,*} \and
Gu\'enol\'e Silvestre \inst{1,6} \and 
Kathleen M. Curran\inst{1,2}\thanks{Corresponding Author}\orcidID{0000-0003-0095-9337}
}
\authorrunning{F. Wang et al.}
%
\institute{
Research Ireland Centre for Research Training in Machine Learning \and
School of Medicine, University College Dublin, Dublin, Ireland \and
The Third People's Hospital of Zhenjiang City, Zhenjiang, China \and
Zhenjiang Maternal and Child Health Hospital, Zhenjiang, China \and
The Fifth People's Hospital of Zhenjiang City, Zhenjiang, China \and
School of Computer Science, University College Dublin, Dublin, Ireland
\email{408418010@qq.com,kathleen.curran@ucd.ie} \\
\inst{\dagger} These authors contributed equally to this work.
}
\maketitle              
\begin{abstract}
Clinician-centered evaluation is critical for validating medical \ac{AI} systems, especially in ultrasound imaging where quantitative metrics do not always capture clinical usability. Existing medical image platforms primarily focus on dataset labeling. They lack integrated support for blinded model comparison and reproducible evaluation workflows. We present a clinician-centered pipeline for remote annotation and evaluation in ultrasound \ac{AI} studies. The proposed pipeline uses a centralized server and lightweight browser interfaces to enable clinicians to perform annotation, blinded ranking, and review without local dataset downloads. The pipeline also supports multi-rater participation, centralized result aggregation, and automated statistical analysis. We validate the pipeline in a fetal ultrasound segmentation study with six raters spanning expert, generalist, and non-expert experience levels. The system automatically generated Spearman correlation, Kendall's $\tau$, and top-1 selection statistics. Results indicated moderate to strong agreement across experts and other groups. The blinded evaluation results showed a tendency for later active learning models to be preferred. These outcomes suggest that the pipeline can support clinician-centered annotation and reproducible human-\ac{AI} evaluation studies in ultrasound imaging. The proposed pipeline is available on \href{https://github.com/13204942/SonoRate}{GitHub}.

\end{abstract}
\section{Introduction}

Deep learning has become widely adopted in medical image analysis for segmentation, classification, detection, and diagnostic support~\cite{Chen:2022,Kim:2022,Rayed:2024,Groh:2024,Aggarwal:2021}. In ultrasound imaging, recent \ac{AI} systems have shown effectiveness for fetal biometrics, anatomical structure segmentation, and disease assessment~\cite{Bano:2021,Zeng:2021,Plotka:2021,Fiorentino:2023,Guo:2024,Plotka:2025,Guo:2026}. Despite these advances, evaluation remains dominated by quantitative metrics such as Dice score, accuracy, and Hausdorff distance. While standardized, these metrics still fail to capture a model's clinical usability in scenarios where image quality is variable and anatomical boundaries are ambiguous~\cite{Zhou:2020,Boumeridja:2025}.

Clinician-centered evaluation and reader studies are therefore increasingly important for understanding how \ac{AI} predictions align with clinical judgment and workflow requirements~\cite{Reinke:2021,Madan:2025}. Existing medical image platforms commonly support dataset creation and annotation, and they lack integrated support for clinician-in-the-loop \ac{AI} validation studies~\cite{cvat,labelstudio,DiazPinto:2022a,DiazPinto:2022b}. Moreover, clinician evaluation protocols are often developed independently for each project, which hinders workflow consistency and reproducibility across research groups and institutions~\cite{Novak:2024,Livingston:2025}.

Practical deployment poses an additional challenge. Medical imaging data are frequently constrained by institutional governance, privacy regulations, and data-sharing policies, which complicates remote multi-center evaluation~\cite{Bell:2024}. These constraints limit clinician accessibility and make large-scale collaborative validation more difficult. Therefore, lightweight and standardized frameworks are needed to enable efficient remote collaboration while keeping medical imaging data centrally hosted without direct distribution.

In this work, we design a clinician-centered \ac{AI} pipeline to support clinician participation throughout annotation and evaluation. The proposed pipeline uses a centralized server and a lightweight web-based interface so clinicians can interact with the system without local software installation or dataset download. It also conducts automated statistical analysis and report generation for reproducible human-\ac{AI} validation studies.
The main contributions of this work are summarized as follows:
\begin{itemize}
    \item A lightweight, clinician-centered pipeline for medical imaging \ac{AI} evaluation studies that integrates annotation and preference-ranking comparison.
    \item A remote evaluation strategy that preserves data governance by keeping raw medical images centrally hosted while enabling collaborative clinician participation.
    \item A standardized workflow for blinded model comparison, multi-rater preference analysis, and reproducible statistical reporting.
    \item A demonstration study in fetal ultrasound involving clinicians with varying expertise levels and active learning model comparisons.
\end{itemize}

\section{Related Work}

\begin{table}
\centering
\caption{Comparison between existing annotation platforms and the proposed pipeline. \cmark: supported, \xmark: not directly supported, \textasciitilde: partial support.}
\label{tab:comparison}
\resizebox{\linewidth}{!}{
\begin{tabular}{lcccc}
\hline
\textbf{Feature} & \textbf{CVAT} & \textbf{Label Studio} & \textbf{MONAI Label} & \textbf{Ours} \\
\hline
Browser-based annotation & \cmark & \cmark & \cmark & \cmark \\
Medical imaging support & \textasciitilde & \textasciitilde & \cmark & \cmark \\
Annotation management & \cmark & \cmark & \cmark & \cmark \\
Remote deployment & \cmark & \cmark & \cmark & \cmark \\
Blinded model comparison & \xmark & \xmark & \xmark & \cmark \\
Preference ranking workflow & \xmark & \xmark & \xmark & \cmark \\
Multi-rater agreement analysis & \xmark & \xmark & \xmark & \cmark \\
Automated statistical reporting & \xmark & \xmark & \xmark & \cmark \\
\hline
\end{tabular}
}
\end{table}

\subsection{Medical Image Annotation Platforms}

Several annotation tools have been developed to support medical image analysis and dataset construction. General annotation tools such as CVAT~\cite{cvat} and Label Studio~\cite{labelstudio} provide flexible interfaces for image labeling, segmentation, and collaborative dataset management. In medical imaging, specialized tools including MONAI Label~\cite{DiazPinto:2022a,DiazPinto:2022b} and ITK-SNAP~\cite{Yushkevich:2006} have further enabled interactive segmentation and AI-assisted annotation workflows. These frameworks have contributed substantially to the development of annotated datasets for training deep learning models. However, existing tools are primarily designed for dataset annotation and labeling rather than clinician-centered \ac{AI} evaluation. They focus on generating \ac{GT} annotations or correcting model predictions, with limited support for blinded comparison studies between multiple \ac{AI} models. In particular, existing tools generally do not provide standardized workflows for clinician ranking, preference analysis, or multi-rater agreement assessment, which are important for evaluating the clinical relevance of models' predictions. Moreover, in medical imaging, data sharing is often restricted by privacy and institutional policies~\cite{Bell:2024}. Existing tools provide limited support for lightweight remote evaluation workflows, which enable collaborative clinician participation without direct dataset distribution.

As shown in Table~\ref{tab:comparison}, existing platforms primarily focus on annotation and dataset creation, whereas our pipeline additionally supports blinded model comparison, clinician preference ranking, multi-rater agreement analysis, and automated statistical reporting for clinician-centered \ac{AI} evaluation studies.

\subsection{Human-Centered \ac{AI} Evaluation}

Human-centered \ac{AI} evaluation has become increasingly important in medical image analysis, where quantitative metrics may not fully reflect clinical usability or diagnostic relevance~\cite{Maier-Hein:2018}. Reader studies provide an important way to assess whether \ac{AI} systems are consistent with clinician interpretation and whether \ac{AI} can support real diagnostic workflows~\cite{Obuchowski:1996,Jassim:2025,Warren:2026,Gommers:2026}. Recent studies have also highlighted the importance of human-AI collaboration and clinician trust for reliable deployment of medical \ac{AI} systems~\cite{Warren:2026,Gommers:2026,Madan:2025}. In particular, preference alignment between clinicians and \ac{AI} models has emerged as an important direction for understanding whether improvements in quantitative performance are consistent with human judgment and clinical expectations~\cite{Maier-Hein:2018,Reinke:2021}. However, existing studies often use evaluation protocols that provide limited support for standardized blinded comparison and reproducible statistical analysis.

\section{Pipeline Design}

\subsection{Overall Pipeline}

\begin{figure}[t]
\centering
    \includegraphics[width=\linewidth]{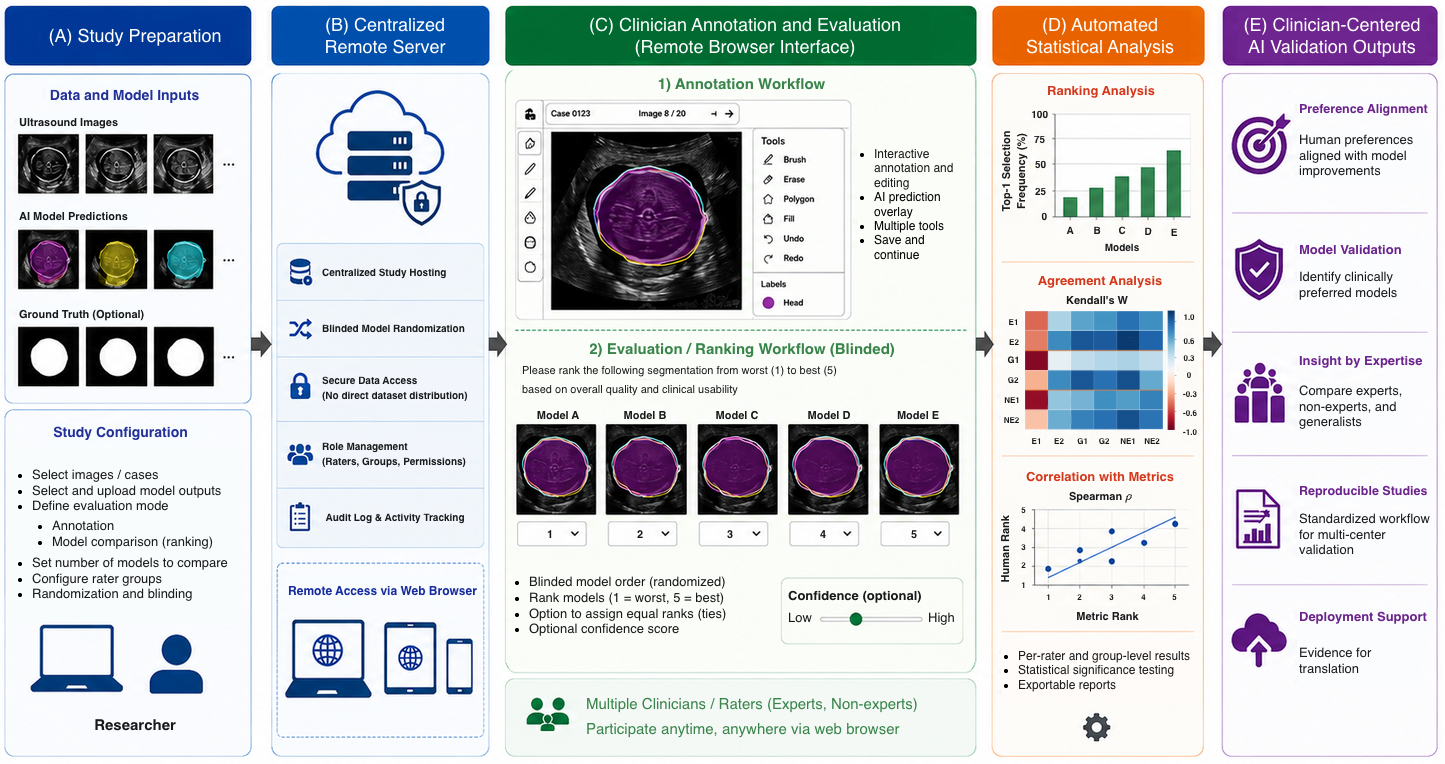}
\caption{Overview of the proposed clinician-centered pipeline for remote annotation and evaluation in ultrasound imaging studies. The framework supports study configuration, browser-based clinician interaction, blinded model comparison, annotation workflows, and automated multi-rater statistical analysis.}
\label{fig:framework}
\end{figure}

An overview of the proposed clinician-centered annotation and evaluation pipeline is presented in Fig.~\ref{fig:framework}. This pipeline consists of the following components: (1) researcher server, (2) clinician client, (3) secure image streaming, (4) annotation and ranking modules, (5) result aggregation, and (6) statistical analysis.

The researcher server is responsible for study preparation, centralized resource management and statistical analysis. Researchers upload ultrasound images, \ac{AI} model predictions, and optional reference annotations to the server. The server also manages study configuration, including evaluation mode, clinician groups, randomized model ordering, and blinded comparison settings. All images and model outputs are hosted centrally to avoid direct distribution of medical datasets.

A lightweight web-based client is used to allow clinicians to interact with the system. This client is executable on the server without installation. Ultrasound images and segmentation overlays are streamed directly from the centralized server to the browser interface. This design enables remote clinicians-in-the-loop while simplifying collaborative evaluation across multiple users and institutions.

The pipeline also supports both annotation and evaluation workflows. In the annotation stage, clinicians can create and edit segmentation masks using an interactive tool~\cite{Dutta:2019}. In the ranking stage, clinicians are presented with multiple anonymized model outputs and asked to rank these outputs based on overall quality and clinical usability. The model outputs are ordered randomly and independently for each clinician to ensure blinded evaluation.

Clinicians' annotations, rankings, and preference scores are collected and saved on the researcher server. The statistical analysis module subsequently computes agreement metrics, such as Spearman correlation, Kendall's $\tau$, top-1 selection frequency, and inter-rater agreement statistics. These results provide a standardized pipeline for clinician-centered validation and human-centered \ac{AI} evaluation studies.

\subsection{Researcher Server and Remote Deployment}

\begin{figure}[t]
\centering
    \includegraphics[width=\linewidth]{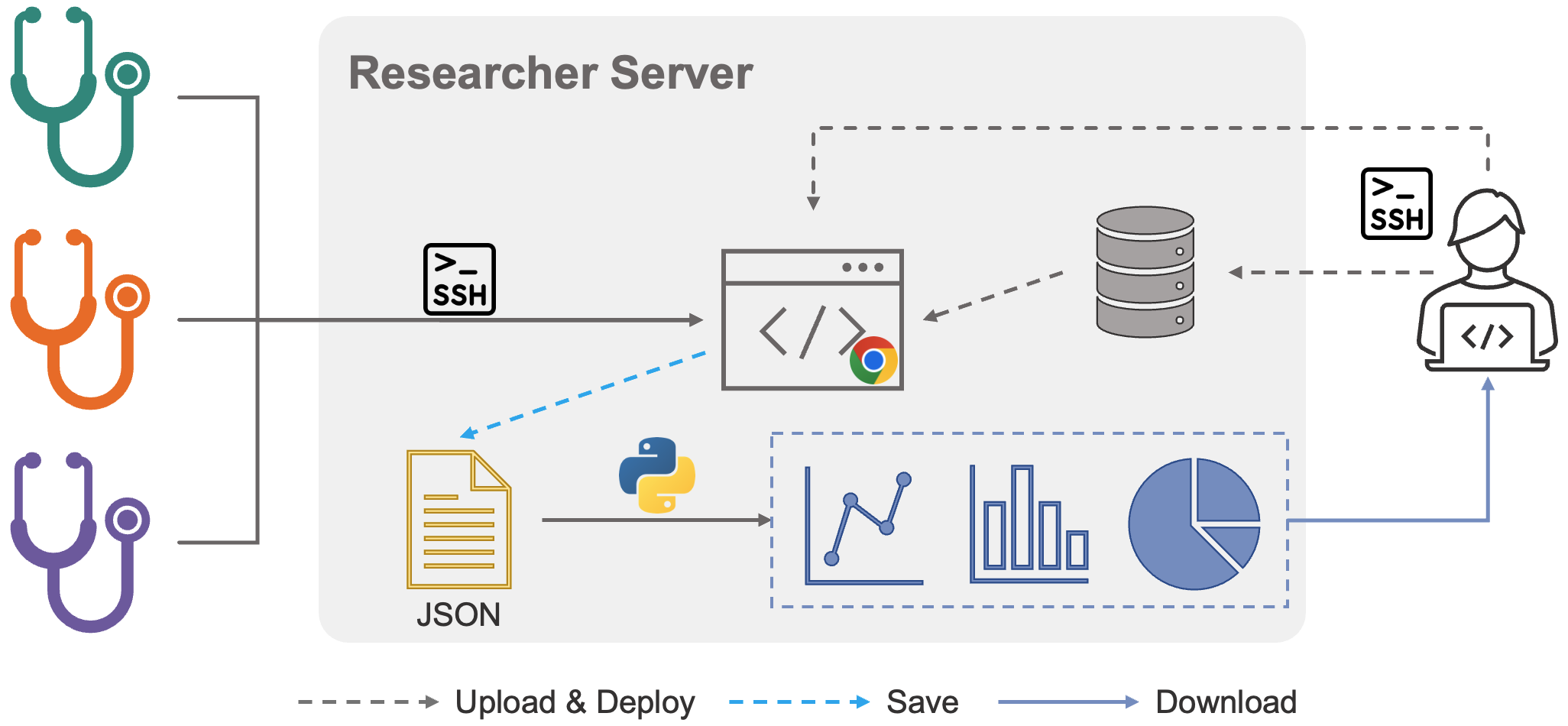}
\caption{Overview of the remote deployment workflow for clinician and researcher collaboration.}
\label{fig:remote_server}
\end{figure}

Fig.~\ref{fig:remote_server} presents the researcher server functionality and remote deployment workflow. The proposed pipeline uses a centralized deployment strategy where all images, \ac{AI} model predictions, annotations, and study settings are hosted on a researcher server. Researchers can upload data, define annotation tasks, configure rater groups, and select evaluation modes through the server. Clinicians access the system remotely using a lightweight browser interface without installing additional software. During interaction, images and segmentation overlays are streamed directly from the server. All clinician annotations and ranking results are automatically saved on the server, while a Python program performs subsequent statistical analysis and report generation.

Researchers manage the deployment process and upload study data to the researcher server through a network connection using the Secure Shell (SSH) protocol. The annotation and evaluation interfaces are lightweight and can be easily modified to support different clinical tasks and study requirements.

\subsection{Annotation and Ranking Interfaces}

\begin{figure}[t]
\centering
    \includegraphics[width=.7\linewidth]{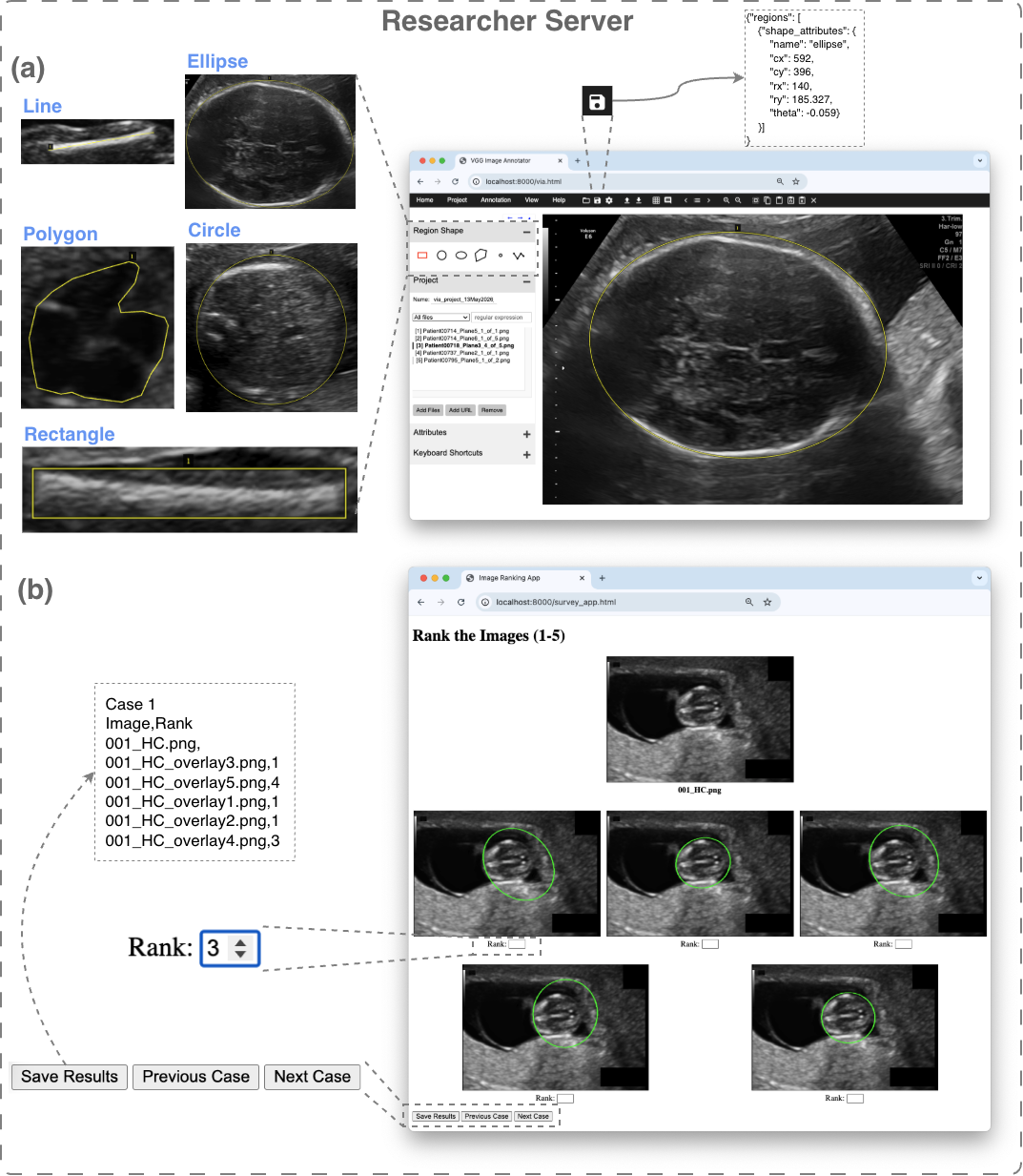}
\caption{The lightweight interface for clinician annotation and evaluation via browser. (a) The annotation interface supports multiple annotation shapes and exports annotations in JSON format. (b) The ranking interface presents ultrasound images with overlaid segmentation masks in randomized order. The ranking results are automatically saved as text files.}
\label{fig:interfaces}
\end{figure}

Fig.~\ref{fig:interfaces} shows screenshots of the annotation and ranking interfaces. The annotation interface supports multiple annotation shapes and interactive editing, as well as JSON-formatted annotations. The ranking interface displays images with overlaid segmentation masks and allows clinicians to rank model outputs using scores from 1 to 5. The ranking score of each case is saved in a plain text file. All annotations and ranking results are automatically saved and aggregated by a Python program running on the researcher server.

\subsection{Result Aggregation and Statistical Analysis}

All clinician annotations and rankings are automatically collected and stored on the researcher server during the study. A Python program subsequently aggregates the results and generates statistical reports for evaluation. The proposed pipeline supports multiple agreement and ranking analyses, including Spearman correlation, Kendall's coefficient, top-1 selection frequency, and inter-rater agreement statistics. These outputs provide a standardized workflow for clinician-centered evaluation and human-AI agreement analysis.

\section{Experimental Demonstration}

\begin{figure}
\centering
    \includegraphics[width=\linewidth]{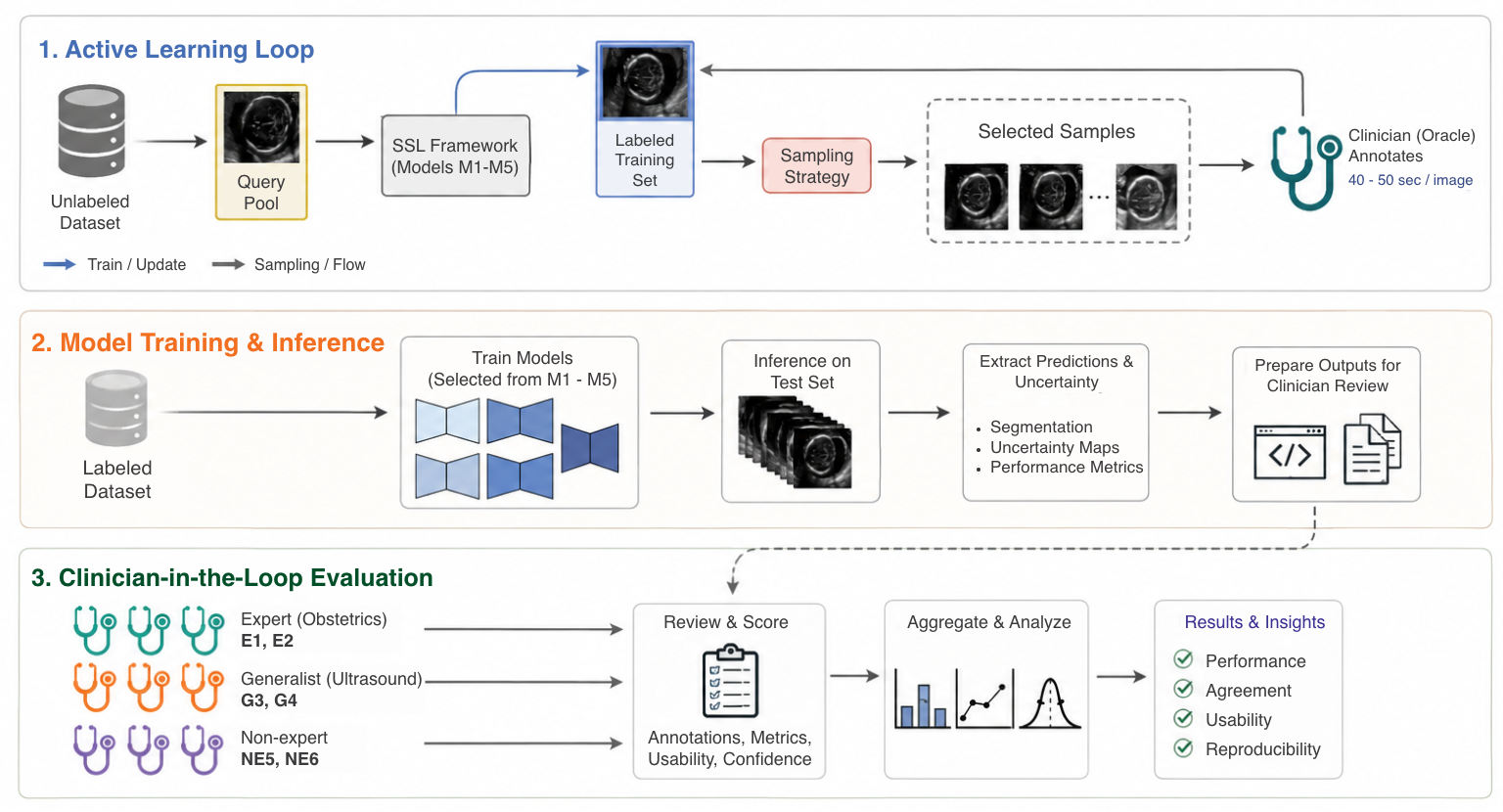}
\caption{Overview of the clinician-centered pipeline for our ultrasound AI studies. The workflow has active learning, sampling strategy, model training and inference, and clinician-in-the-loop evaluation.}
\label{fig:example_workflow}
\end{figure}

This section demonstrates the clinician-centered pipeline in a fetal ultrasound study. We used two public fetal ultrasound datasets, HC18 and ES-TCB. HC18 contains fetal head ultrasound images acquired during routine obstetric examinations~\cite{Heuvel:2018}, while ES-TCB contains trans-cerebellum fetal ultrasound images collected in Spain~\cite{Xavier:2020,Alzubaidi:2023}. Building on a semi-supervised learning framework~\cite{Wang:2025c}, we developed an active learning pipeline and evaluated five segmentation models (M1--M5) from different active learning iterations via the proposed clinician-centered pipeline, as shown in Fig.~\ref{fig:example_workflow}.

A total of six raters participated remotely through the browser interface. There were two expert obstetric sonographers (E1, E2), two general ultrasound sonographers (GE3, GE4), and two non-expert participants (NE5, NE6). The details of these raters are presented in Table~\ref{tab:raters}. The pipeline enabled all raters to complete the study without local installation or direct dataset download. 
The browser-based interface enabled efficient remote participation. On average, clinicians required approximately 45--50 seconds to annotate an ultrasound image and around 30 seconds per case for blinded ranking. These observations suggest that the proposed pipeline can support practical clinician participation with relatively low interaction overhead.
For each dataset, 30 cases were randomly selected for rating, resulting in 60 cases in total. Each case contained five anonymized model predictions randomly presented for blinded comparison. In the end, this study collected 300 model rankings from each rater.

\begin{table}[t]
\centering
\caption{Details of the raters recruited for the human evaluation study.}
\label{tab:raters}
\begin{tabular}{ccccc}
\hline
\textbf{Rater ID} & \textbf{Role} & \textbf{Specialty} & \textbf{Experience (Y)} & \textbf{Clinical Group} \\
\hline
E1 & Sonographer & Obstetrics & 15 & Specialist \\
E2 & Sonographer & Obstetrics & 5 & Specialist \\
G3 & Sonographer & Generalist Ultrasound & 18 & Generalist \\
G4 & Sonographer & Generalist Ultrasound & 18 & Generalist \\
NE5 & Non-expert & N/A & 0 & Non-expert \\
NE6 & Non-expert & N/A & 0 & Non-expert \\
\hline
\end{tabular}
\end{table}

\begin{figure}[tb]
\centering
    \includegraphics[width=\linewidth]{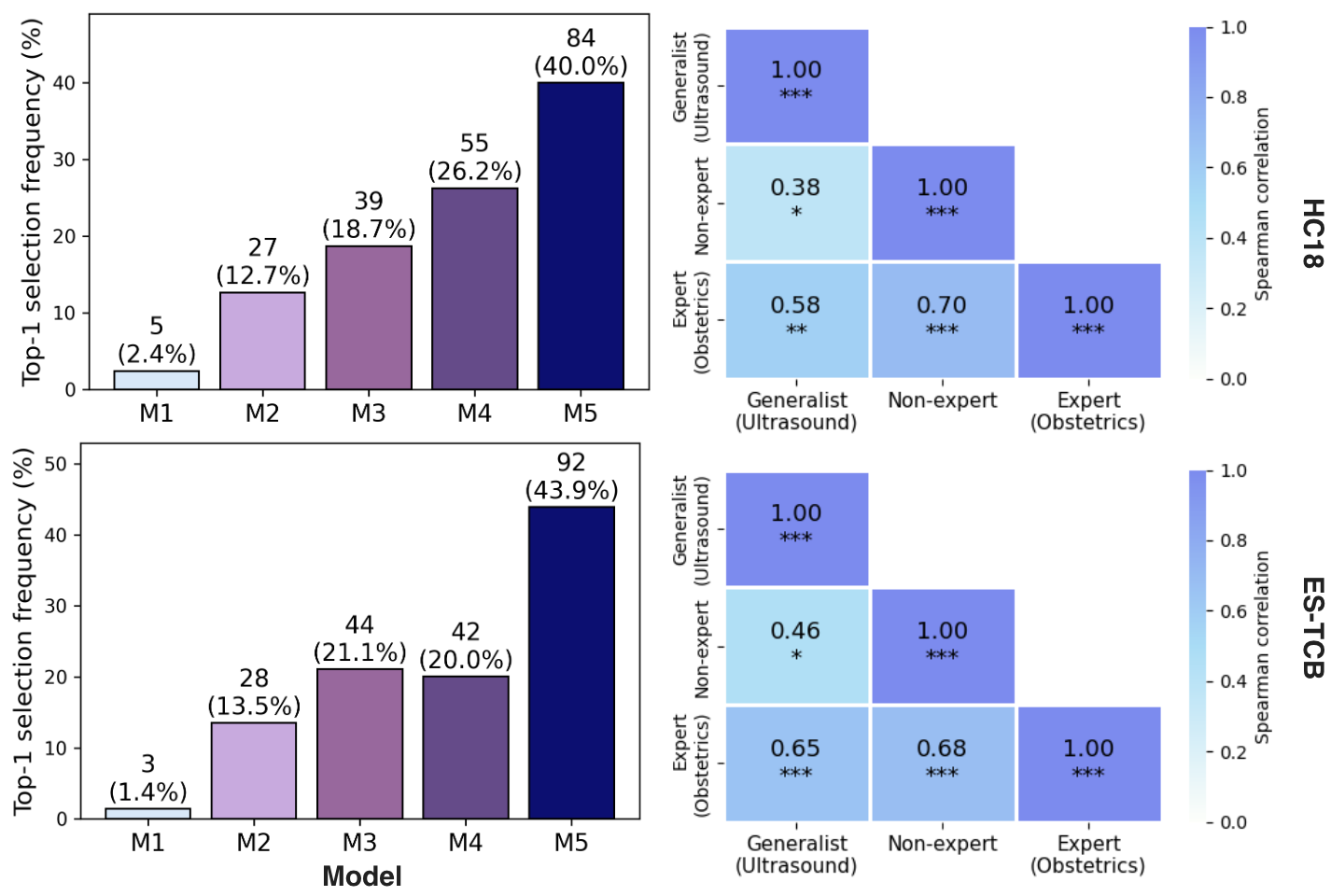}
\caption{Top-1 selection frequency across models (M1--M5) and inter-group agreement measured by Spearman correlation among experts, generalists, and non-experts on HC18 and ES-TCB datasets. ***: $p<0.001$. **: $p<0.01$. *: $p<0.05$.}
\label{fig:selection}
\end{figure}

Fig.~\ref{fig:selection} presents an example of the statistical analysis generated by the proposed pipeline in our demonstration experiment. In both the HC18 and ES-TCB datasets, later models (M3-M5) were selected more frequently during blinded evaluation, while earlier models (e.g., M1) were rarely preferred. The pipeline also supports inter-group agreement analysis across expert, generalist, and non-expert raters using Spearman correlation. Moderate to strong positive correlations were observed across different rater groups, indicating consistent preference patterns during remote clinician evaluation.

Table~\ref{tab:kendall_w} shows an example of inter-rater agreement analysis using Kendall’s $\tau$. Moderate agreement was observed across both HC18 and ES-TCB datasets, with mean agreement values of 0.53 and 0.55, respectively. The pipeline also reports variability statistics, including standard deviation and interquartile range, which indicate differences in agreement across evaluation cases. A limited proportion of images achieved high inter-rater agreement ($\tau > 0.7$), accounting for 20.0\% in HC18 and 26.7\% in ES-TCB. Lower agreement was observed in challenging cases characterized by ambiguous anatomical boundaries, poor ultrasound image quality, or minimal visual differences between model predictions.

\begin{table}
\centering
\caption{Inter-rater agreement (Kendall’s $\tau$) across datasets.}
\setlength{\tabcolsep}{4.5pt}
\begin{tabular}{lcccccccc}
\hline
Dataset & Mean & Median & Std & IQR & Min & Max & $\tau>0.7$ (\%) & $\tau<0.4$ (\%) \\
\hline
HC18   & 0.53 & 0.51 & 0.18 & 0.15 & 0.12 & 0.87 & 20.0 & 16.7 \\
ES-TCB & 0.55 & 0.56 & 0.22 & 0.21 & 0.12 & 0.85 & 26.7 & 20.0 \\
\hline
\end{tabular}
\label{tab:kendall_w}
\end{table}

\section{Discussion}



This paper introduced a clinician-centered pipeline for remote annotation and evaluation in ultrasound \ac{AI} studies. The pipeline integrates clinician participation, blinded model comparison, annotation workflows, and automated statistical analysis within a unified web-based interface. Unlike existing annotation platforms that primarily focus on dataset labeling, our design supports clinician-in-the-loop \ac{AI} validation and reproducible multi-rater evaluation studies while preserving data governance through centralized hosting.

The pipeline offers translational value for medical \ac{AI} validation studies. By enabling remote clinician participation, it facilitates reader studies and multi-center validation experiments without requiring clinicians to download raw datasets. The integrated statistical analysis module provides consistent reporting of agreement metrics and ranking outcomes, making the pipeline suitable for preference alignment studies and clinical usability assessments.

Although a formal usability study was outside the scope of this work, the demonstration experiment showed that clinicians were able to complete annotation and ranking tasks efficiently through the browser interface. The average annotation time was approximately 45--50 seconds per image, while blinded ranking required approximately 30 seconds per image.

Several limitations remain. The current implementation assumes pre-established collaboration between researchers and clinicians. Therefore, the implementation does not yet include enterprise-level authentication, role-based access control, or audit logging. Moreover, the additional usability measures, such as clinician satisfaction, perceived workload, and user experience, are not investigated in this work. Lastly, the current implementation is optimized for ultrasound imaging and has not yet been extensively evaluated across other modalities. 

Future work will first incorporate secure user authentication, permission management, and enhanced governance features to support large-scale multi-center clinical studies better. Then we will investigate additional usability measures to improve clinician experience. Afterward, we will extend support to additional medical imaging modalities. 

\section{Conclusion}

We presented a secure, clinician-centered, end-to-end evaluation pipeline that enables reproducible remote human-\ac{AI} annotation and validation studies in ultrasound imaging without direct medical data sharing. The proposed pipeline is lightweight and easily deployable for efficient clinician participation, blinded model comparison, multi-rater evaluation, and statistical analysis. A demonstration study showed that the proposed pipeline enables clinicians to annotate fetal head segmentation labels, supports six raters in conducting human-\ac{AI} evaluation, and performs agreement analysis without direct dataset distribution. We hope this work can facilitate collaborative clinician-in-the-loop studies and support the development of more clinically aligned medical \ac{AI} systems.

\begin{credits}
\subsubsection{\ackname} This work was funded by Taighde \'{E}ireann – Research Ireland through the Research Ireland Centre for Research Training in Machine Learning \\
(18/CRT/6183).

\subsubsection{\discintname}
The authors have no competing interests to declare that are relevant to the content of this article. 
\end{credits}


%
%
%
\bibliographystyle{splncs04}
\bibliography{refs}
%




\end{document}